\documentclass[proceedings, preprint]{rmaa}

\usepackage{paralist}

\usepackage{psfrag,color}

\SetYear{2008}
\SetConfTitle{MFUII}

\title{ Magnetic Fields and the Polarization of Astrophysical Maser Radiation:A Review}

\author{William D. Watson \altaffilmark{1}}
\altaffiltext{1}{Department of Physics, University of Illinois,1110 West Green Street,
Urbana, IL 61801-3080 USA(w-watson@uiuc.edu).}

\shortauthor{Watson}
\shorttitle{Magnetic Fields and Maser Polarization}
\listofauthors{W.D. Watson}
\indexauthor{Watson, W.D.}

\abstract{Basic aspects of the relationship between the magnetic field and polarized
maser radiation are described with the emphasis on interpreting the observed
spectra. Special attention is given to three issues---the limitations on the
applicability of the classic solutions of Goldreich, Keeley \& Kwan (1973),
inferring the strength of the magnetic field from the circular polarization
when the Zeeman splitting is much less than the spectral linebreadth
(especially for SiO masers), and the significance of the absence of
components of the Zeeman triplet in the spectra of OH masers in regions of
star formation.}

\addkeyword{circumstellar matter}
\addkeyword{ISM:clouds}
\addkeyword{magnetic fields}
\addkeyword{masers}
\addkeyword{polarization}

\begin{document}
\maketitle

\section{INTRODUCTION}

The extreme surface brightnesses and the narrow spectral line
breadths of astrophysical masers make them ideal sources for high precision
observational studies with very long baseline interferometry, as well as in
studies with a single dish. Strong masing was first detected in the
mid-1960's--- in the 18 cm transitions of the OH molecule. Strong masing was
subsequently detected in the transitions of H$_{2}$O, SiO and CH$_{3}$OH.
Masing occurs in star-forming regions, in the circumstellar gas of evolved
stars, in the gaseous disks at the nuclei of active galaxies, and in distant
galaxies. The radiation often is polarized so that conclusions can be drawn
about the direction and strength of the magnetic fields.

One might think that information about the magnetic fields in masers would not be
generally useful since the density, temperature,and history of the gas in
masers is so different from that of the bulk of the gas of interstellar clouds.   
However, there is evidence that this is not the case. The directions of ordered
magnetic fields of the OH masers in star forming regions are similar to the
directions of the field in the larger volume surrounding the masers (Fish et
al. 2003). The field strengths inferred for the OH and H$_{2}$O masers in
star forming regions lie on the same curve of magnetic field $B\propto $
(density)$^{1/2}$ as those obtained for the interstellar gas where the
density is lower by many orders of magnitude (Fiebig \& Gusten 1989). The
patterns in the directions of the linear polarization of circumstellar
masers strongly suggests that they reflect the direction of a larger scale
magnetic field (Kemball \ et al. 2008) and the variation of the inferred
field strengths with distance from the star has been interpreted as the
variation of a large scale field (Vlemmings 2007). Upper limits to the field
strength in the masing disk around the massive black hole at the nucleus of
the galaxy NGC4258 have been used to restrict the mass accretion rate and
the nature of the accretion process (Modjaz et al. 2005). Magnetic field
strengths in distant galaxies have also been inferred from the circular
polarization of maser radiation (Robishaw et al. 2008).

In the review here, I will discuss issues involved in extracting information
about the magnetic field from the observed polarization characteristics of
astrophysical maser radiation. After a general orientation, I will focus
mainly on three topics: limitations in applying the classic work of
Goldreich, Keeley \& Kwan (1973; designated here as GKK) to the observations, the
validity of using the Zeeman effect to infer strong magnetic fields from
the circular polarization of SiO masers, and the significance of the characteristics
of the Zeeman components of OH maser spectra for turbulence in star forming
regions. More comprehensive reviews of the observations include those of Lo
(2005) and of Vlemmings (2007).

\section{GENERAL CONSIDERATIONS}

\subsection{Basic Ideas}

Astrophysical masers are considered as tubular with the
consequence that the radiation is beamed, perhaps highly beamed. The solid
angle $\Delta \Omega $ into which the radiation is beamed is quite
uncertain. Unless the stimulated emission rate $R$ due to the maser
radiation is negligible in comparison with the loss rate $\Gamma $ of the
molecular states due to all other processes, the maser radiation must be
included in the calculation of the molecular populations and the radiative
transfer calculation is thus nonlinear. The ratio $R/$ $\Gamma $ is then the
degree of saturation. Even when the angular size of the masing feature can
be resolved and the surface brightness is determined, $R/$ $\Gamma $ is
still uncertain because $\Delta \Omega $ cannot be measured. Estimates can
be made. From these it seems that $R/$ $\Gamma $ probably is less than about
10, but is not much less than 1---typically, at least for prominent masers.

Almost all theoretical discussions of maser polarization in the
literature (a) consider the maser geometry to be completely linear, (b)
consider only the two energy states of the masing transition (these are
split into substates by interaction with the magnetic field), and (c)
consider only maser transitions between molecular states of low angular
momentum (esp. J=1-0 transitions). The premise is, of course, that the
results are indicative for actual astrophysical masers.

With the above estimates for $R$ and the strengths of the magnetic
fields that seem to be present, it is almost always (if not always) the case
that the Zeeman splitting $g\Omega $ $\gg R,\Gamma $ where $g\Omega $ is the
splitting in ordinary frequency units multiplied by 2$\pi $. It follows that
the magnetic field direction is an axis of symmetry and the ordinary
magnetic substates are ``good quantum states''. A key simplification is then
valid: (d) the calculation can be performed with radiative transfer
equations that are of the same form as those for ordinary thermal spectral
lines and the populations of the magnetic substates can be found by solving
ordinary rate equations. Otherwise, it would be necessary to use the much
more involved formulation in terms of quantum mechanical density matrices as
done by GKK. We are fortunate. If $g\Omega $ $\gg R$ were not satisfied, it
is unlikely that useful information about the magnetic field could be
inferred from the polarization data. The behavior of the polarization
becomes quite complicated when $R$ approaches and exceeds $g\Omega ,$ as can
be seen from density matrix calculations for this regime ( Nedoluha \&
Watson 1990,1994). Because the magnetic field provides an axis of symmetry
when $g\Omega $ $\gg R$, the direction of the linear polarization of the
radiation emitted by the molecules must be either parallel or perpendicular
to the direction of the magnetic field as projected on the sky. The
direction of the linear polarization that we observe may be altered from
this by propagation effects (especially Faraday rotation), though in
practice such effects do not seem to be important except possibly for the 18
cm OH masers.

Conveniently, astrophysical masers divide into two classes according
to whether the Zeeman splitting is much weaker (SiO, CH$_{3}$OH and H$_{2}$O
masers) or much stronger (mainline OH masers) than the spectral linebreadth.

\subsection{The Formulation of Goldreich, Keeley and Kwan (GKK)}

GKK commonly has served as the starting point for calculations and
discussions about the polarization of astrophysical maser radiation---and
rightly so!\ It is thus important to be clear about what GKK did and did not
do, and what are the restrictions on the expressions that they obtain as
solutions to their equations.\ GKK consider a linear maser permeated by a
constant magnetic field and involving an angular momentum J=1-0 radiative
transition.\ The pumping and the loss of excitation due to causes other than
the maser transition are not specifically treated, but are expressed as
``phenomenological'' pump $\Lambda $ and loss $\Gamma $ rates ---a
proceedure that often is used in laboratory as well as in astrophysical
maser theory. GKK obtain solutions only in certain limits---including the limits of strong and
weak splitting. I will discuss the solutions of GKK for weak splitting since
this is where most of the confusion has arisen in recent years. GKK consider
only the center of the spectral line in obtaining solutions for weak
splitting. Understanding the GKK equations requires considerable effort
because they are expressed in terms of quantum mechanical density matrices
for the magnetic substates of the masing molecules. GKK do this so that the
equations can be used when $R\approx g\Omega $ and $R\gg g\Omega $, as well
as when $R\ll g\Omega $ which we now realize to be the regime of chief (if
not exclusive) importance for astrophysical masers.

The key point about the GKK\ solutions to their equations is that
the solutions are obtained in certain asymptotic limits of the degree of saturation $R/\Gamma $
---though still subject to certain other restrictions that define various
regimes. In the regime $R\ll g\Omega $ which is of practical interest and when $R/\Gamma $ $\gg 1$, 
GKK obtain the simple formula for the fractional linear polarization (Stokes $%
Q/I $) at line center that is probably the most quoted result of their
paper,
\begin{eqnarray*}
Q/I &=& (3\sin ^{2}\theta -2)/3\sin ^{2}\theta  for  \sin
^{2}\theta \geq 1/3 \\
and &=&-1 for \sin ^{2}\theta \leq 1/3.
\end{eqnarray*}
where $\theta $ is the angle between the propagation direction and the
magnetic field.

This is a valuable result, but \textbf{it should not be compared directly
with the observations} because the $R/\Gamma $ for astrophysical masers are
not so large that solutions for $R/\Gamma \gg 1$ are a good
approximation---as we have shown in a number of papers by numerically
integrating the GKK equations to provide solutions for $Q/I$ as a function
of $R/\Gamma $ as shown in Figure 1.

\begin{figure*}[!t]
  \includegraphics[width=\textwidth,height=20cm]{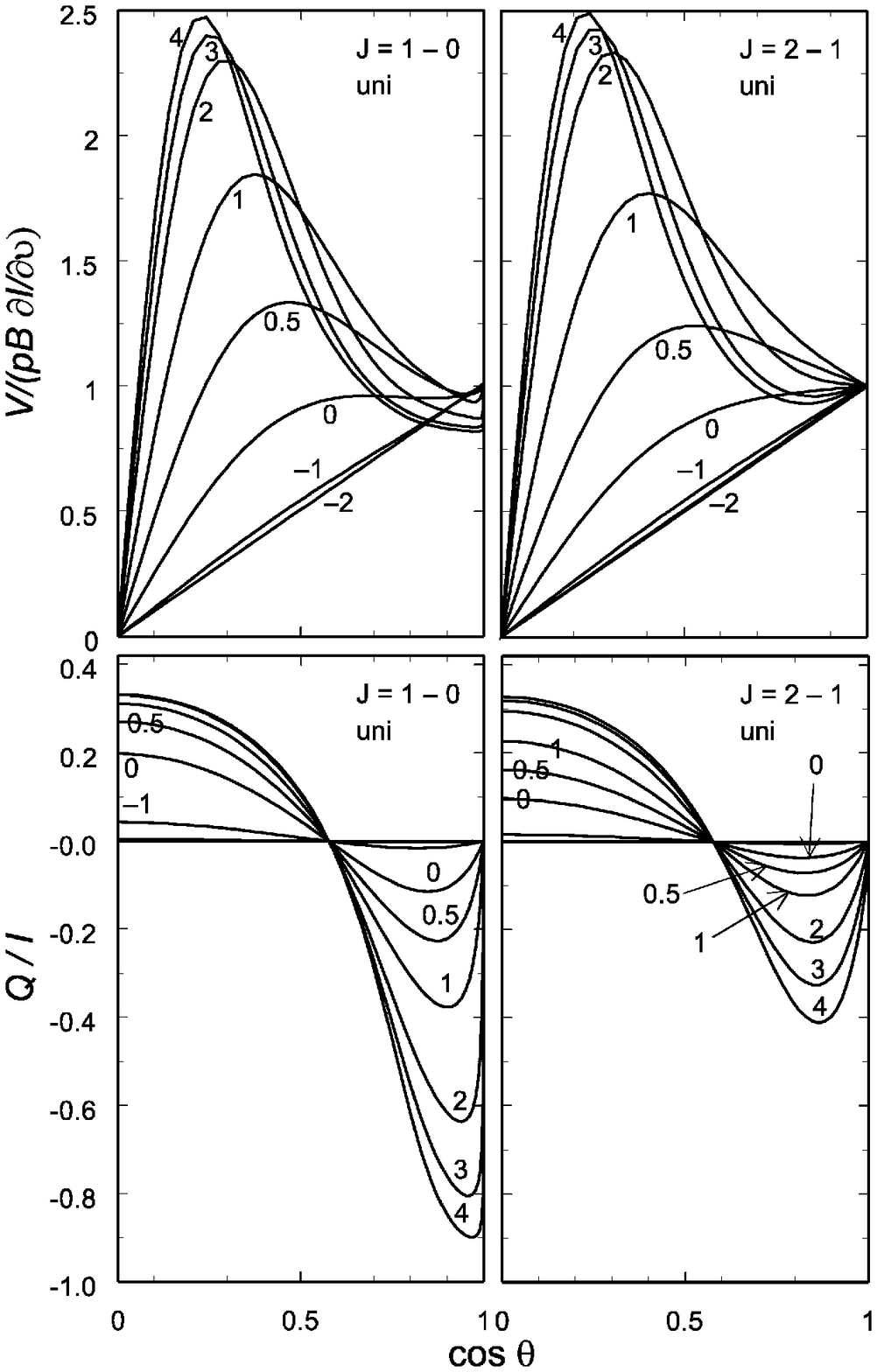}%

\caption{Circular and linear polarization of maser radiation as a
function of the cosine of the angle $\theta$ between the direction
of the magnetic field and the line of sight. Polarizations are
presented for angular momentum $J=1-0$
and $J=2-1$ masers, in separate panels as
indicated by the label in
each panel.The curves are labeled
by the $\log_{10}$ of the saturation degree. \\
(upper three panels) The circular polarization is measured by the
magnitude of
$V/(pB\partial I/\partial v)$ which is equal to $\cos\theta$ in the unsaturated
limit.\\
(lower three panels) The fractional linear polarization. Stokes-$Q$
is positive when the linear polarization is perpendicular to the direction
of the
magnetic field projected onto the plane of the sky. (from Watson \& Wyld 2001)}

  \label{fig:f1mod.eps}
\end{figure*}

That this GKK solution is not directly relevant can also be seen
simply by comparing with the observations. The prominent H$_{2}$O and CH$%
_{3} $OH masers typically exhibit fractional linear polarizations less than
a few percent and the weakly split OH masers, no more than about ten
percent. Such low polarization is obtained in the above GKK solutions only
at an angle $\theta $ of almost exactly 55 degrees. It is clearly
implausible that the radiation from all such masers propagates at only this
angle relative to the magnetic field! Reasonable estimates for the beaming
angles $\Delta \Omega $ together with observed brightnesses also indicate
that $R/\Gamma $ is not so large that GKK solutions are directly applicable.
The spectral line widths and profiles of maser radiation provide additional 
evidence that $R/\Gamma $ is not
large ( Nedoluha \& Watson 1991;Watson, Sarma, \& Singleton 2002 )

It is true that one prominent class of masers with weak Zeeman splitting
--- the circumstellar SiO masers --- does exhibit high fractional linear
polarization. However, the problem with these is that the linear polarization is
too high! Too high, at least for plausible $R/\Gamma $ (see subsection 3.2).

\textbf{To summarize}, the GKK solution in the weak splitting regime
is not directly applicable to the observations mainly because (a) it is only
for asymptotically large $R/\Gamma $. Other limitations are that GKK (b)
consider only isotropic pumping and (c) consider only a J=1-0 transition.
The limitation (b) is critical for the SiO masers (subsection 3.2). Any electric
dipole transition has the asymptotic $Q/I$ of the GKK solution. However,
with increasing angular momentum of the molecular states in the transition,
larger and larger values of $R/\Gamma $ are required for the actual
(numerical) solution to approach the GKK solution as indicated in Figure
1. Finally, (d) in the weak splitting regime, GKK consider only line center 
and thus have nothing to say
about the circular polarization. The circular polarization is always zero at
line center in this idealization.

\section{THE WEAK SPLITTING REGIME AND SiO MASER POLARIZATION}

\subsection{Masers Other than SiO Masers}

The GKK solution reproduced above demonstrates the key feature of intense ($%
R/\Gamma \gtrsim 1$), beamed radiation in the presence of a magnetic field
in the weak splitting regime when $g\Omega $ $\gg R$---that the intense
radiation will depopulate the magnetic substates in a way that leads to
polarization of the beam itself. Though the $R/\Gamma $ for astrophysical
masers is not large enough that the GKK solution for the linear polarization
is quantitatively accurate, the GKK solution does accurately tell us the
direction of the linear polarization as a function of the angle of
propagation of the maser beam including the angle of 55 degrees (the ``van
Vleck angle'' ) at which the linear polarization changes abruptly from
parallel to perpendicular relative to the projected direction of the
magnetic field---as long as the pumping is isotropic.

The fractional linear polarizations observed for the H$_{2}$O and CH$_{3}$OH
masers typically are no more than a few percent and are consistent with the
low polarizations expected from the numerical solutions to the GKK equations
shown in Figure 1 for masers with little or no radiative saturation.
As mentioned in the previous Section, estimates for $R/\Gamma $ are consistent with the
low values implied by the fractional polarization.  For these transitions
which involve molecular states with higher angular momentum, the calculated
fractional polarization at a specific $R/\Gamma $ also tends to be lower
than in Figure 1 (see Nedoluha \& Watson 1990,1992). The weakly split OH
masers associated with supernova shells exhibit somewhat higher fractional
linear polarizations ($Q/I\approx \ 10$ percent) which would be consistent
with modest saturation ($R/\Gamma \approx \ 1$) for the low angular momentum
of the states\ involved and a plausibly wide range of possible propagation
angles $\theta $.

For these masing species, the fractional circular polarization that is
detected, together with the Zeeman interpretation, leads to magnetic field
strengths that seem to be ``reasonable''. That is, the field strength lies
near $B\propto $ (density)$^{1/2}$ when this relationship is extrapolated
from lower densities and the magnetic pressures are similar to the gas
thermal pressures. For the H$_{2}$O and CH$_{3}$OH masers, the low
fractional linear polarization and the low degree of saturation are
additional factors that tend to make it difficult for non-Zeeman causes to
be responsible for the circular polarization. Hence, the Zeeman
interpretation is expected to be at least approximately correct for the
circular polarization of these masers.

However, the circular polarization of even the ordinary Zeeman effect can be
altered somewhat by maser saturation as shown in Figure 1. Although GKK do
not consider frequencies away from line center in the weak splitting regime
where the circular polarization would be non-zero, their equations can be
generalized in a straightforward manner. The results of the numerical
integration of these generalized GKK equations is what is shown in Figure 1.
When $R/\Gamma \lesssim 0.1$, the relationship between $B$ and Stokes $V$
differs negligibly from that for ordinary thermal lines ($V \propto B\cos
\theta $) as seen in Figure 1. When saturation is significant, the
relationship changes so that $V$ can actually increase rather than decrease
as $\theta $ increases---a variation first demonstrated in the context of
the polarization of water masers (Nedoluha \& Watson 1992). \ Nevertheless,
$R/\Gamma $  is unlikely to be large enough for any of these masers to cause
more than ``factor of two'' uncertainties in the magnetic fields that are
inferred by using the standard relationship for thermal lines and omitting
the uncertain $\cos \theta $ factor.

\subsection{SiO masers}

Understanding the polarization characteristics of the circumstellar SiO
masers is more challenging. In the following, I will continue to assume that
$g\Omega $ $\gg R$ is satisfied as is indicated from ``best estimates''. It
is somewhat less clear, however, that this inequality is so strongly
satisfied as for the other types of masers. The magnetic moment is somewhat
less than for H$_{2}$O and the spectral line profiles of the SiO masers are
less amenable for drawing conclusions about the degree of saturation. For
the magnetic moment of the SiO transition, $g\Omega =$ $1.5B$(mG).
Noticeable deviations from the ``$g\Omega $ $\gg R$ polarizations'' begin to
appear when $R\simeq g\Omega /10$ (Nedoluha \& Watson 1990,1994). Of course,
if the magnetic fields are as strong as inferred from the Zeeman
interpretation of the observed circular polarizations, $g\Omega $ $\gg R$
should be well satisfied.

$Linear$ $polarization.$

Observations have shown for many years (e.g., Clark, Troland, \& Johnson
1982) that the fractional linear polarization of the prominent J=2-1 (v=1)
masing transition can be 50\% and greater; it can also be as large for the
weaker and less well studied masing transitions involving states of higher
angular momentum or higher vibration. While such high polarization is a
feature of the GKK solution, it would require that the degree of saturation
be implausibly large. A saturation degree $R/\Gamma >10^{4}$ would be
required according to calculations in Figure 1. The natural resolution is
that the polarization is at least partly a result of the pumping. That is,
the magnetic substates are excited unequally because the angular
distribution of the infrared radiation involved in the pumping (mainly that
due to vibrational v=1-0 transitions) is anisotropic. This can happen if the
direct infrared radiation from the star is important in exciting the v=1
states from the v=0 ground states or if the optical depths for the escape of
the v=1-0 radiation emitted by the SiO molecules following excitation by
collisions are anisotropic. Detailed calculations have been performed to
show that such anisotropic pumping is likely (Deguchi \& Iguchi 1976\
;Western \& Watson 1983; Ramos et al. 2005). 

Depending upon the details of
the anisotropic pumping, the angle at which the change in sign occurs for $Q$
will be different from the 55 degrees of Figure 1. In fact, the sign of $Q$
can be the same at all angles. The approximately 90 degree difference in the
direction of the linear polarization that is sometimes observed for SiO
masers at nearby locations can be due to changes in the directions of the
anisotropy relative to the magnetic field and the line of sight (Western \&
Watson 1983; Ramos et al. 2005).\ Since $g\Omega $ $\gg R$ probably is
satisfied, the direction of the linear polarization will be either parallel
or perpendicular to the projected direction of the magnetic field---as long
as it is unaffected by propagation effects. The observed patterns seem to
indicate that such propagation effects are not a major factor for
circumstellar SiO masers.

$\ Circular$ $polarization.$

The original detections of the circular polarization of the circumstellar
SiO masers (Barvainis, McIntosh \& Predmore 1987) and the interpretation
that this polarization is a straightforward consequence of the Zeeman effect
yield magnetic field strengths of 10-100G! Reasonable estimates can be made
for the kinetic temperature and the maximum gas density of the masing gas.
With these and $B=$ 10G, the magnetic pressure exceeds the thermal gas
pressure by a factor of 1000! This may indicate the presence of interesting
magnetic phenomena in the circumstellar envelopes, but the implication of
such unusual conditions also suggests that the Zeeman interpretation of the
circular polarization should be severely scrutinized beyond the ``factor of
two'' corrections that potentially follow simply from radiative saturation
as given in Figure 1.

Although a cause of non-Zeeman circular polarization that must be kept in
mind is the possibility that $g\Omega $ $\gg R$ is not sufficiently well
satisfied (Nedoluha \& Watson 1994), this does not now seem to be the most
likely possibility. Magnetic fields of some 100mG are detected from the
water masers, and those at the SiO masers are expected to be at least this
large since the SiO masers seem to be somewhat closer to the star. A more
likely possibility is that a misalignment occurs within the masing gas
between the directions of the linear polarization of the maser radiation and
the projected direction of the magnetic field as the radiation propagates
through the maser. Linearly polarized radiation will then be converted to
elliptically polarized in much the same way as is well known in classical
optics when linearly polarized radiation passes through a plate in which the
phase velocity is different along the two orthogonal, optical axes (e.g.,
the ``quarter wave plate''). The phase velocity near an absorption or
emission feature depends on the strength of the feature and changes rapidly
with frequency. We know that the strength of the emission feature is
different along two orthogonal axes in the masing gas since it is just this
difference that is creating the strong linear polarization of the observed
maser radiation and, of course, the maser radiation itself is linearly polarized
within the masing gas. We do not, however, know whether the third requirement for
this non-Zeeman circular polarization is satisfied---that the optical axes
and the direction oof the linear polarization become misaligned within the
masing region. The circumstellar medium appears quite irregular (due perhaps
to convection, turbulence and/or shock waves) and it seems plausible that
the magnetic field can change direction somewhat and cause the required
misalignment. Faraday rotation within the masing gas would have a similar
effect. The Stokes V that would result from this non Zeeman cause would also
be antisymmetric about line center because of the antisymmetric variation of
the phase velocity. It can thus easily be confused with the antisymmetric
profile of Stokes V due to the Zeeman effect.

What is the observational evidence? The SiO masers are believed to be
somewhat closer to the stars and stronger fields are expected than for the H$%
_{2}$O masers where the fields are more reliably determined to be in the hundreds of milliGauss. 
It can be reasoned that much stronger fields of 10G or so at the
location of the SiO masers represent a plausible extrapolation of the
weaker fields at larger distances from the stars (Vlemmings 2007), though
this is a large extrapolation. On the other hand, in single dish
observations there seems to be a correlation between the fractional linear
polarization and the fractional circular polarization of SiO maser radiation
from star to star based on the observation of a large number of stars
(Herpin et al. 2006). Since circular polarization in non-Zeeman processes is
created from a small fraction of the linear polarization, a correlation
tends to support a non Zeeman origin. However, the significance of single
dish observations, which can be summing over numerous masing components
around each star, is unclear. The fractional linear polarization in the weak
splitting regime is independent of the strength of the magnetic field.
Hence, there should be no correlation between the linear and circular
polarizations if the variation in the circular polarization is due to
variations in the magnetic field strength and the Zeeman effect.  Because
both polarizations can increase with angle and with saturation in Figure 1,
an apparent correlation might occur if the range of propagation angles in
Figure 1 is limited because of, perhaps, preferred viewing angles. Strong
masing in both the J=1-0 and J=2-1 transitions of v=1 is widely observed and
a clear prediction can be made for the ratio of their fractional circular
polarizations based on the Zeeman effect---assuming that the degree of
saturation either is small or is equal for the two transitions when the
relationship in Figure 1 is applied. To be sure that the radiation of the
two transitions is from the same masing environment, VLBI with good velocity
resolution is required to separate the numerous masing spots in the
circumstellar environments. The results of such studies are not yet
available.

\section{The Strong Splitting Regime and MHD Turbulence in Regions of Star
Formation}

Soon after the discovery of astrophysical masers, it was recognized that the
spectra of these OH 18 cm masers from regions of star formation tended to be
highly circularly polarized. A small fraction of the features can be
identified as pairs of Zeeman sigma components from the same spatial
locations. The components are well separated in frequency due to Zeeman
splitting and the magnetic field strength can then be reliably inferred.
Except in one region, the Zeeman pi components do not seem to be present.
The comprehensive study of \ Garcia-Barreto et al. (1988) established the
statistics of the polarization properties of the numerous masing features
that tend be present at various locations in a cloud. Though the most
extensive data come from the 18 cm transition, the absence of pi components
also seems to be a feature of the analogous masing transitions of the excited
states of OH in regions of star formation (e.g., Caswell \& Vaile 1995). I
will thus accept as the observational characteristics to be understood: (a)
only a single Zeeman component is usually observed and (b) though pairs of
sigmas are sometimes detected, the pi components are essentially never
present. This behavior is almost certainly a result of irregularities\
(turbulence, waves, etc.) in the velocities and magnetic fields of the gas.

The strong tendency for only a single Zeeman component is the easier to
understand---at least conceptually. If there are gradients in both the
magnetic field and the velocity of the gas along the path of the radiation,
the change in the magnetic field will cause the rest frequency of one of the
Zeeman components to vary in the sense that tends to compensate for the
apparent change in the frequency of the radiation due to the varying Doppler
shift (Cook 1966). The maser optical depth will then be largest for one of
the sigma components. In addition, if the velocity differences are large
enough that the maser radiation\ from one Zeeman transition ``sweeps
through'' the frequencies of other Zeeman transitions at other locations as
it traverses masing gas, a single sigma component can also emerge (Deguchi
\& Watson 1986) even without a variation in the strength of the magnetic
field along the path of the radiation. Detailed calculations demonstrate
that the spectral line can remain narrow under these conditions as observed
(Nedoluha \& Watson 1990b). It is evident that this process can also mix the
polarization characteristics of \ Zeeman components in the radiation at a
specific observed frequency.

The absence of the pi components is more challenging. The presence of pairs
of Zeeman sigmas indicates that velocity and magnetic field gradients are
not always sufficient to reduce the spectrum to a single component. However,
the angular distributions of pi and sigma radiation are different---with the
peak maser optical depths for the sigmas being along the magnetic field and
the peak for the pi components, perpendicular to the magnetic field. Gray
and Field (1995) emphasized that there may be additional considerations that
favor the propagation of maser radiation at small angles to the field lines
so that we tend to observe only at propagation angles where the sigma
components are favored. Because of the ``exponential gain'' associated with
masers, the intensities of the pi components will be negligible at the
smaller angles where their maser optical depths are smaller, consistent with
the observations.

A reason for the OH maser optical depths to be greater in directions near
the magnetic field lines may be found in  the properties of MHD turbulence.
Coherent regions in the turbulence tend to be elongated along the lines of
the magnetic field (Goldreich \& Shridar 1995). Hence the gradients in
velocity tend to be smaller along the field lines, and the maser optical
depths greater, than in other directions. Calculations have been performed
that demonstrate the plausibility of this idea at a more quantitative level
by calculating the maser optical depths at various angles to the mean
magnetic field using the results of MHD simulations to represent the
turbulence in the masing gas (Watson et al. 2004; also Wiebe \& Watson
2006). The degree of anisotropy in MHD turbulence depends upon the ratio of
the Alfven velocity to the sound speed. The calculations indicate that this
ratio must be at least 3 to understand the absence of the pi components of
these OH masers.

\section{SUMMARY}

Prevailing estimates for the strengths of the magnetic fields in the
environments of astrophysical masers indicate that essentially always the
Zeeman splitting in frequency units is so much greater than the rates for
other processes (stimulated emission, etc.) that ``ordinary methods'' can be
used in calculations for the polarized maser radiation.

The classic paper of GKK provides us with the methods to perform the
calculations in all regimes, but their widely cited result for the
fractional linear polarization of masers with weak Zeeman splitting is not
quantitatively applicable because it is obtained for maser saturation that
is much higher than actually occurs. The formulation of GKK must also be
extended to include anisotropic pumping (which seems to be essential for the
SiO masers) and to calculate the circular polarization in the weak splitting
regime (needed to infer field strengths). Neither is treated by GKK.

Whether the circular polarization of the SiO masers is due to the Zeeman
effect, and hence whether the inferred magnetic fields of 10G or so actually
are present in these circumstellar environments should be considered as 
uncertain at present.

The tendency for only one sigma component of the Zeeman triplet of OH masers
to be detected in regions of star formation and (apparently) for the pi
components to be absent is almost certainly related to irregularities in the
velocities and magnetic fields. Calculations of the masing in which
simulations of MHD turbulence are used to describe these irregularities lead
to spectra that are\ generally similar to what is observed. In this
interpretation, the absence of pi components is related to the anisotropic
nature of MHD turbulence.

\center REFERENCES
\begin{description}
\item{Caswell, J.L., \& Vaile, R.A. 1995, MNRAS, 273,328}
\item{Clark, F. O., Troland, T. H., \&\ Johnson, D. R. 1982, ApJ, 261, 569}
\item{Cook, A.H. 1966, Nature, 211,503}
\item{Deguchi, S., \& Iguchi, T., 1976, PASJ, 28,307}
\item{Deguchi, S., \& Watson, W.D. 1986, ApJ, 300, L15}
\item{Fiebig, D., \& Gusten, R. 1989, A\&A, 214, 333}
\item{Fish, V.L., \& Reid, M.J., Argon,\ A.L., \&\ Menten, K.M. 2003, ApJ, 506, 328}
\item{Goldreich, P., Keeley, D.,\& Kwan, J.Y. 1973, ApJ, 179, 111}
\item{Goldreich, P., \& Shridar, S. 1995, ApJ, 438, 763}
\item{Gray, M.D., \& Field, D. 1995, A\&A, 298, 243}
\item{Herpin, F., Baudry, A., Thum, C., Morris, D., \& Wiesemeyer, H. 2006, A\&A,450, 667}
\item{Kemball, A.J., Diamond, P.J., Gionidakis, I., Mitra, M., Yim, K., \& Pan, K.-C., 2008, in preparation}
\item{Lo, K.Y. 2005, ARA\&A, 43,625}
\item{Modjaz, M., Moran, J. M., Kondratko, P. T., \& Greenhill, L. J. 2005, ApJ,626, 104}
\item{Nedoluha, G.E., \& Watson, W.D.\ 1990a, ApJ, 354,660}
\item{Nedoluha, G.E., \& Watson, W.D.\ 1990b, ApJ, 361,653}
\item{Nedoluha, G.E., \& Watson, W.D.\ 1991, ApJ, 367, L63}
\item{Nedoluha, G.E., \& Watson, W.D.\ 1992, ApJ, 384,185}
\item{Nedoluha, G.E., \& Watson, W.D.\ 1994, ApJ, 423,394}
\item{Ramos, A.A., degl'Innocenti, E.L., \& Bueno, J.T. 2005, ApJ, 625, 985}
\item{Robishaw,T., Quataert, E., \& Heiles, C. 2008, ApJ, in press}
\item{Vlemmings, W.H.T. 2007, arXiv 0705.0885}
\item{Watson, W.D., Sarma, A. P., \& Singleton, M.S. 2002, ApJ, 570, L37}
\item{Watson, W.D., Wiebe, D. S., McKinney, J.C., \& Gammie, C.F. 2006, ApJ, 604,707}
\item{Watson, W.D., \& Wyld, H. W. 2001, ApJ, 558, L55}
\item{Western, L.R., \& Watson, W.D. 1983, ApJ, 275, 195}
\item{Wiebe, D. S., \& Watson, W.D. 1998, ApJ, 503, L71}
\item{Wiebe, D., S. \& Watson, W.D. 2007, ApJ, 655, 275}
\end{description}

\end{document}